\documentclass{iaus}
\usepackage{graphicx}

\title[Constraints on the Formation and Evolution of Globular 
Cluster Systems] %% give here short title %%
{Observational Constraints on the Formation and Evolution of Globular 
Cluster Systems}

\author[Stephen E. Zepf]   %% give here short author list %%
{Stephen E. Zepf}

\affiliation{Dept. of Physics and Astronomy, Michigan State University, 
East Lansing, MI 48824, USA \break e-mail: zepf@pa.msu.edu}

\pubyear{2008}
\volume{246}  %% insert here IAU Symposium No.
\pagerange{}
\date{}
\jname{Dynamical Evolution of Dense Stellar Systems}
\editors{E. Vesperini, M. Giersz, A. Sills}
\begin{document}
\maketitle

\begin{abstract}
This paper reviews some of the observational properties
of globular cluster systems, with a particular focus on those that 
constrain and inform models of the formation and dynamical evolution 
of globular cluster systems. I first discuss the observational
determination of  the globular cluster luminosity and mass function.
I show results from new very deep HST data on the M87 globular
cluster system, and discuss how these constrain models of
evaporation and the dynamical evolution of globular
clusters.
The second subject of this review is the question of how
to account for the observed constancy of the globular cluster
mass function with distance from the center of the host
galaxy. The problem is that a radial
trend is expected for isotropic cluster orbits, and while
the orbits are observed to be roughly isotropic, no
radial trend in the globular cluster system is observed.
I review three extant proposals to account for this, and
discuss observations and calculations that might determine
which of these is most correct.
The final subject is the origin of the very weak mass-radius
relation observed for globular clusters. I discuss how this
strongly constrains how globular clusters form and evolve.
I also note that the only viable current proposal to account for
the observed weak mass-radius relation naturally affects the
globular cluster mass function, and that these two problems
may be related.

\end{abstract}

%\section % if your document starts with a section,
              % remove some space above using thi

%\section{Introduction}

%The focus of this review is on observational studies of globular
%cluster systems 

\section{What is the Shape of the Globular Cluster Luminosity Function?}

A natural starting point for determining the key physical
processes that make the globular cluster luminosity function (GCLF)
is to consider the best available observational constraints on the 
GCLF. The ideal observational sample for determining the GCLF
would both very deep to constrain the faint, low-mass end of 
the cluster population, and have large numbers of clusters
to provide adequate statistics, particularly for the rare
very bright and very faint globular clusters.
The Milky Way globular cluster system meets the first criterion
of probing to very faint globular clusters, and
has thus provided some of the basic data indicating the
roughly lognormal shape of the GCLF, and the location of
the turnover magnitude and estimated mass of the distribution.
However, the Milky Way has a relatively
poor globular cluster system, numbering only about 150
objects. As a result the Galactic GCLF provides very limited 
statistical power for any test of the behavior of the GCLF,
particularly at the low and high mass ends, each of which
are only expected to have a few clusters given the total
number of Galactic globulars.

One way to overcome the problem of small numbers of globular 
clusters in the Milky Way is to study a galaxy with a 
much richer system of globular clusters. M87 is an obvious choice
as it has a combination of an exceedingly rich globular cluster
system (with almost two orders of magnitude more globular
clusters than the Milky Way), and a moderate distance with 
its location in the Virgo cluster. However, because the
distance to M87 is not negligible, to realize its potential
for the study of the low-mass end of the GCLF requires data
that reach to faint apparent magnitudes. Moreover, because
the number of background galaxies rises steeply
to faint magnitudes, the data must have excellent 
spatial resolution to distinguish between faint compact galaxies
and M87 globular clusters. At brighter magnitudes, ground-based
imaging with various image classification and color cuts can
create useful globular cluster samples (e.g.\ Rhode \& Zepf 2001
for a detailed example).
However, given the depths required to reach the faint end of the
M87 GCLF, and the declining numbers of globular clusters and rising
number of galaxies at these faint magnitudes, deep
HST imaging is required to reliably probe the shape of the 
faint end of the GCLF in M87 and other similar galaxies.

\begin{figure}
% \vspace*{-2.0 cm}
\begin{center}
 \includegraphics[width=3.4in]{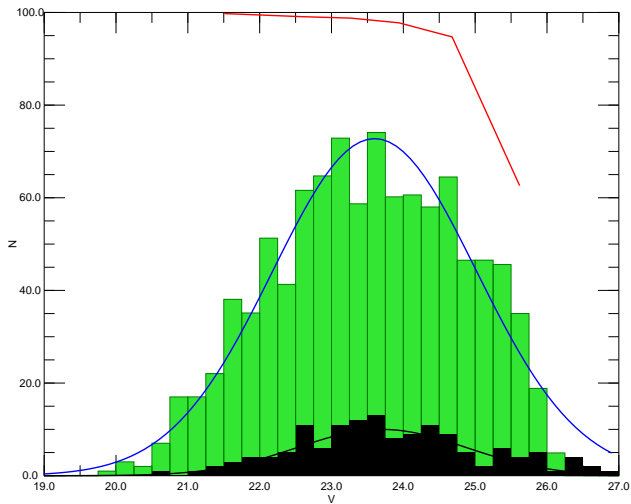}
% \vspace*{-1.0 cm}
 \caption{A plot of the globular cluster luminosity function for 
M87 (upper histogram in green) and the Milky Way (lower histogram in black), 
with the Milky Way data shifted to the 16 Mpc
distance of Virgo. The M87 data is from Waters et al.\ (2006) and
has been completeness corrected, using the completeness function 
shown as the upper (red) line. Gaussian fits to the M87 and Milky Way
data are overplotted on the histograms. The plot makes immediately
apparent the dramatic advantage of the much larger numbers of
globular clusters in M87 for studying the effects of dynamical
evolution on globular cluster systems. As detailed in Waters et al.\ (2006).
the turnover magnitudes of the M87 and Milky Way GCLFs are identical
to within the uncertainties, but the M87 GCLF is significantly broader.
These conclusions are confirmed by the even deeper 50-orbit ACS dataset
(Waters \& Zepf 2008).}
   \label{fig1}
\end{center}
\end{figure}

A start in this direction was made by various multi-orbit HST
WFPC2 studies (e.g.\ Kundu et al.\ 1999). These were able
to establish that the peaks of the turnover of the GCLF were
very similar in M87 and the Milky Way, and also
suggested that the width of the M87 GCLF might be larger
than that of the Milky Way. However, to accurately test
the shape of the GCLF to very faint
and low-mass globular clusters required yet deeper data.
Until recently, such very deep HST data to probe the faint
end of the GCLF were not available.
This situation has changed due to extraordinarily deep imaging 
of M87 obtained originally for studying microlesning in the Virgo
cluster. Two such datasets now exist. The first is a 30 orbit
WFPC2 dataset, for which we have published an analysis
of the faint end of the GCLF in Waters et al. (2006).
The second is a 50 orbit ACS dataset which we are now
analyzing.
In Figure 1, we show the GCLF resulting from our analysis
of the 30-orbit WFPC2 dataset, presented in Waters et al. (2006).
The plot makes immediately
apparent the dramatic advantage of the combination of very 
deep HST imaging and the large numbers of
globular clusters in M87 compared to the Milky Way for studying
the GCLF.

The very deep HST data on the M87 GCLF such as shown in Figure 1
from Waters et al.\ (2006) provide two key constraints on models 
of the dynamical evolution of globular clusters through evaporation.
This can be readily seen when writing the mass loss of a globular
cluster due to evaporation as $\dot{M} = k M^{\gamma}$. $k$ is
then the mass loss rate for a cluster of given mass $M$, and
the exponent $\gamma$ accounts for any dependence of this
mass loss rate on cluster mass. In terms of comparison to
the data, the mass loss rate is primarily constrained by
the location of the turnover of the globular cluster mass 
function, and the dependence of the mass loss rate on $M$
is constrained by the slope of the globular cluster mass
function at low masses. 
These constraints are mostly insensitive
to the initial globular cluster mass function for all extant
models of two-body relaxation and evaporation. Clusters
more massive than the turnover have experienced very little
mass loss as a fraction of their total mass, and the low-mass
end of the GCLF is composed of globular clusters which have lost an increasing
fraction of their initial mass as one goes to fainter globular clusters,
Thus the massive end of the observed GCLF is completely
determined by the initial GCLF (and possibly dynamical
friction for very centrally located clusters) and the low mass 
end of the observed GCLF is set almost entirely by mass loss
from dynamical evolution. As an aside, we note that the
shape around the turnover can be affected in detail
by the initial mass function if this function is flat or 
decreases to lower masses.

\begin{figure}
% \vspace*{-2.0 cm}
\begin{center}
 \includegraphics[width=3.4in]{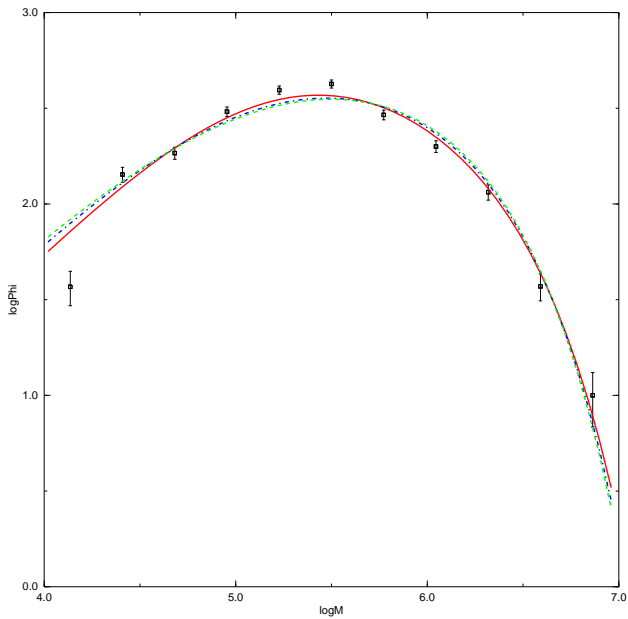}
% \vspace*{-1.0 cm}
 \caption{The mass function of the M87 globular cluster system from
Waters et al.\ (2006) analysis of a 30-orbit HST dataset.
The three lines are models with different dependencies of the
evaporation mass loss rate on cluster mass. The solid line is
$\dot{M} \propto M^{0}$, the dot dashed line is $\dot{M} \propto M^{0.25}$,
and the dashed line is $\dot{M} \propto M^{0.38}$. The data plotted
favor a mass loss rate that is independent of mass, as do
new, even deeper data (Waters \& Zepf 2008).
This result is driven the by the slope from the turnover to 
the low-mass end of the mass function, and is not dependent 
on the initial globular cluster mass function. However, for 
completeness we adopted a Burkert \& Smith (2000) initial mass
function for this plot in order to match the bright end of the 
globular cluster system.}
   \label{fig2}
\end{center}
\end{figure}

We can then compare different theoretical models of mass loss
due to two-body relaxation in globular clusters 
to our extraordinarily deep HST data on the M87 globular cluster
system. Figure 2 plots the results of the comparison
of our M87 GCLF from the 30-orbit WFPC2 dataset 
to several theoretical models as shown in Waters et al.\ (2006).
In detail, three theoretical models are shown, each 
having a different dependence of the mass loss rate on
mass as advocated in recent papers.
One of these possibilities
is that $\dot{M} =$ a constant (i.e. $\gamma = 0$), 
and that the mass loss rate is independent of mass (e.g.\ Fall \& Zhang 2001 
and references therein). Another
possibility comes from the detailed N-body simulations of Baumgardt \&
Makino (2003, hereafter BM03) for which the mass loss rate is
approximately $\dot{M} \propto M^{0.25}$. 
Finally, Lamers, Gieles, \& Portegies Zwart (2005) 
found $\dot{M} \propto M^{0.38}$ from a
combination of fits of the Baumgardt \& Makino (2003) simulations 
and analytic arguments. Additionally, Figure 6
in BM03 indicates that the correct mass to use in this calculation is
the current mass of the globular cluster, and that in the BM03
simulations the mass loss rate of a globular cluster is not
constant in time, and is described well by the $\dot{M} \propto M^{0.25}$
expression calculated for the current cluster mass. There
was some discusson of this point at the meeting, but Figure 6 
from BM03 gives a mass loss rate
due to evaporation that changes as the globular cluster mass
evolves.

The comparison of the deep M87 data to models in Figure 2 favors
mass-loss which is independent of cluster mass. This may be
somewhat surprising in that the most state of the art simulations
would seem to suggest some mass dependence.
In Waters et al. (2006), we discuss some of the possible
resolutions of these differences. One is simply the statistical
difference, even in the very deep 30-orbit WFPC2 data, is not
overwhelming. A second issue noted in Waters et al.\ (2006)
is that the mass-to-light ratio (M/L) of the globular clusters
may decrease as low-mass stars are preferentially lost from
the cluster as it begins evaporating away. We are carrying out
new studies that address these points. Our upcoming work uses 
a 50-orbit ACS
dataset to both reach fainter magnitudes and lower masses and
to increase the overall number. We also use the mass-to-light
ratio evolution of a globular cluster as a function of its 
disruption time given in BM03 to specifically account for the
the changing M/L. The final results from this new work will
be published in Waters \& Zepf (2008).

\section{The Unsolved Problem of the Constancy of the GCLF}

Globular clusters closer to the center of a galaxy will have
smaller tidal radii for a given cluster mass. This will
cause globular clusters closer to the center of a galaxy 
to lose mass more quickly, and therefore a radial gradient 
in the globular cluster mass function and GCLF is expected.
However, the Galactic GCLF shows no strong changes with radius.
This disrepancy between basic theoretical expectation and observation
has been noted for some time (e.g.\ Baumgardt 1988, Vesperini 1987 and
references therein). One way around this problem is to have the
globular cluster orbits become increasingly radial with distance 
from the center of the host galaxy, so that all globular clusters
have very similar pericenters. Coupled with the assumption that
the tidal radii are set at pericenter, one can then recover a
constant GCLF with distance from the center of the host galaxy.
This scenario was investigated in detail by Fall \& Zhang (2001)
who suggested it could account for the lack of an observed
radial gradient in GCLFs. 

The question is whether the existing globular cluster population
has such extremely radial orbits. M87 provides an ideal place to test
proposals for the physical origin of the radial constancy
of the GCLF. As shown in Vesperini et al.\ (2003), HST
data over a very wide range of radii, from 1 to 75 kpc, indicate a 
generally constant GCLF with galactocentric distance. 
\begin{figure}
% \vspace*{-2.0 cm}
\begin{center}
\includegraphics[width=2.3in]{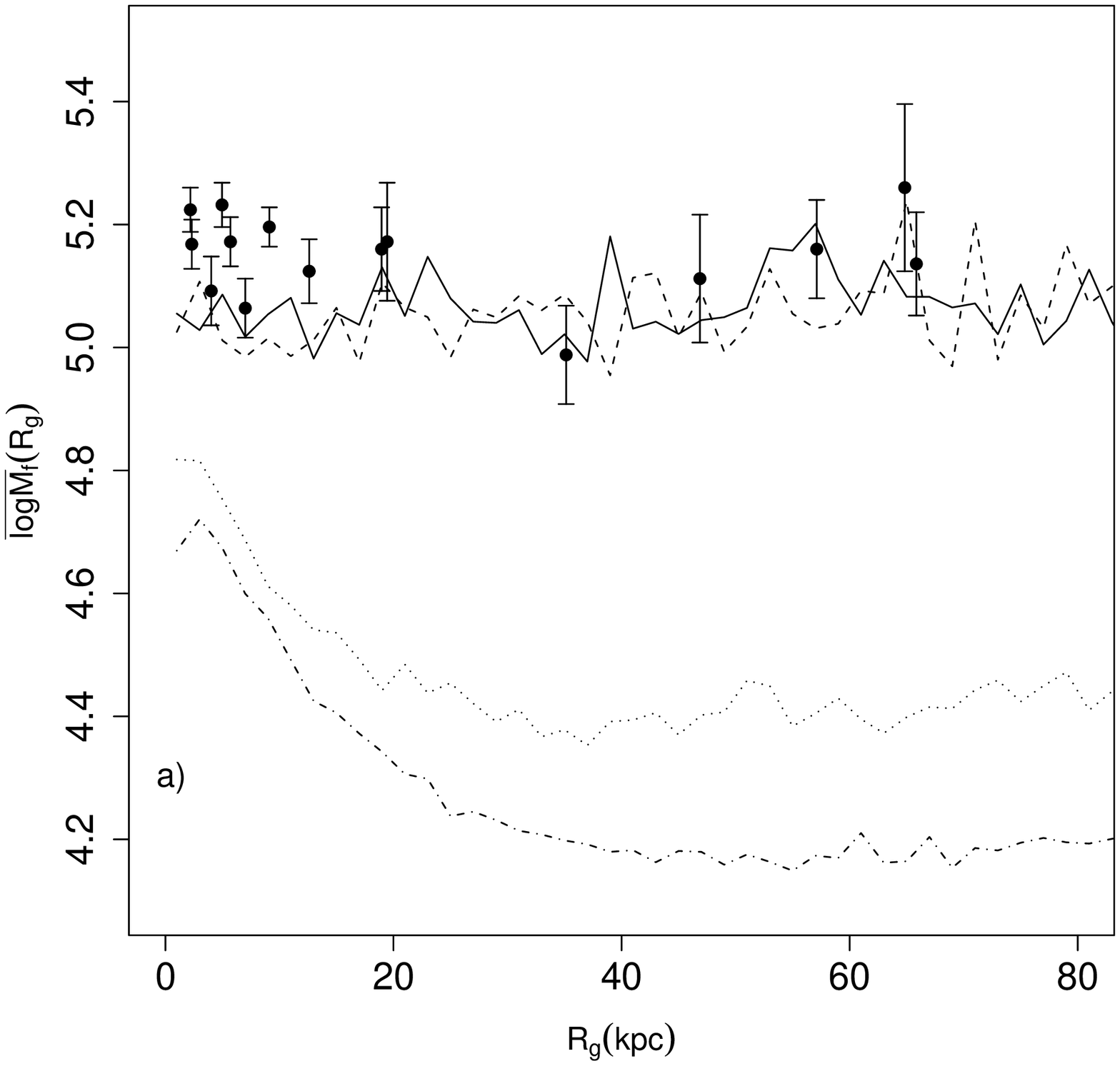}
\includegraphics[width=2.3in]{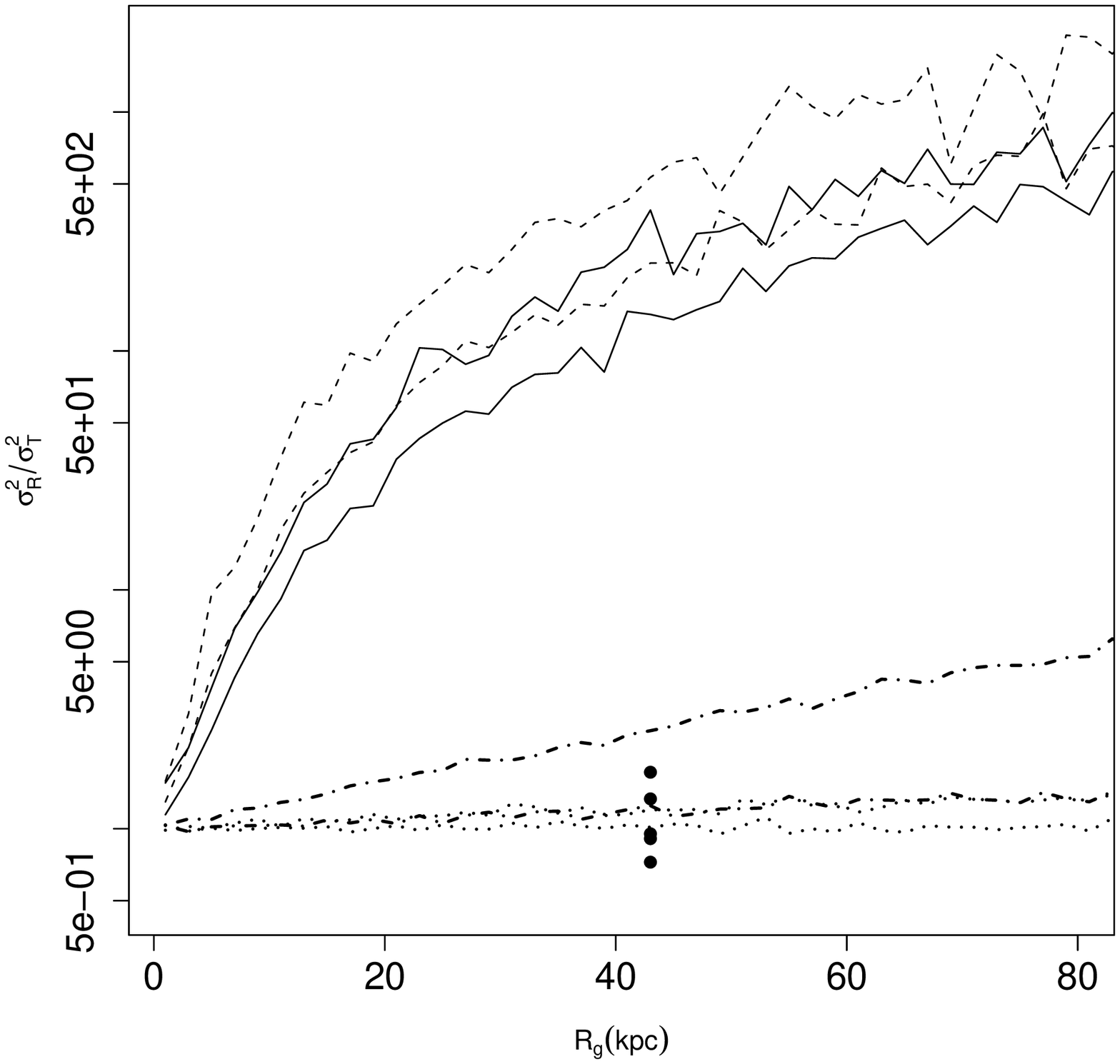}
% \vspace*{-1.0 cm}
 \caption{The plot on the left-hand side shows the final globular 
cluster mean mass vs. projected galactocentric distance for 
given an initial power-law globular cluster mass function 
and an initial anisotropy radius equal to 2 (solid line), 
3 (dashed line), 15 (dotted line), and 150 kpc (dot-dashed line), 
from detailed calculations in Vesperini et al.\ (2003). 
Large anisotropy radii models are close 
to isotropic and small anisotropy radii models have orbits which become 
strongly radial at large galactocentric distances. The constancy
of the GCLF with radius is clearly a problem for isotropic orbits.
The plot on the right compares the anisotropy of the globular cluster
orbits, using the same convention for the models in the plot on the
left, to the observed anisotropy of the M87 globular 
cluster system, which is shown as the points and taken 
from Romanowsky \& Kochanek 2001). This shows that strongly
radial orbits are clearly inconsistent with the observed
velocities. Thus, the dilemma that for a power-law initial
mass function, evaporation-driven dissolution only produces
a constant GCLF with galactocentric distance if the orbits
becomes very radial with distance from the galaxy center,
but the observed radial velocities rule out such radial
orbits (see Vesperini et al.\ 2003 for details).}
\label{fig3}
\end{center}
\end{figure}
Other available
HST data are consistent with the result, but none yet match 
this depth and very large radial range. The other key component
of the M87 globular cluster system is that several radial velocities 
studies of large numbers of M87 globular clusters indicate that the 
current M87 globular cluster population does not have strongly radial 
orbits (e.g.\ C\^ot\'e et al.\ 2001, Romanowsky \& Kochanek 2001). 
Vesperini et al.\ (2003) show 
directly that their observations of a generally constant GCLF with 
distance from the galaxy center combined with the results from the 
radial velocity surveys rule out any model which appeals to strongly
radial orbits in the outer regions of galaxies to explain the
constancy of the GCLF with galaxy radius (see also Figure 3).

What is the explanation for the
generally constant GCLF seen across the Milky Way and M87?
There are three extant proposals, each with its own problems
and promise. In order discussed below, these proposals are - 
1) that hierarchical merging continually mixes the cluster
orbits so thoroughly that clusters now located at a wide
range of distances from the center of a galaxy, i.e. 2 to 100 kpc,
have on average had the same local galaxy potential during
their lifetime as mentioned in Fall \& Zhang (2001), 
2) that early gas expulsion from forming clusters
preferentially disrupts lower-mass proto-globular clusters,
producing a turnover in the GCLF everywhere at early times (e.g.\
Baumgardt, Kroupa, \& Parmentier 2008, and references therein),
and 3) that globular clusters are formed with an initially
power-law mass function, and have a cluster mass-concentration 
relation such that lower mass clusters have lower concentration 
and thus are much more likely to be disrupted by stellar mass
loss (Vesperini \& Zepf 2003).
The difference between 2) and 3) is a matter of physical
origin and timescale, as 3) occurs on a later timescale
in a purely stellar cluster due to stellar mass loss and
smaller concentrations for lower-mass clusters, while 2) happens
more quickly in a proto-cluster which is still gas-rich.

The appeal of the hierarchical model is that it can work
from a power-law initial cluster mass function, as likely
found in the Antennae young cluster system (Zhang \& Fall 1999) 
and consistent with many other young cluster system data.
The problem with this proposal is that the complete mixing
required to make the GCLF today the same at 2 kpc as 100 kpc
has never been demonstrated in any hierarchical model. 
Mergers are generally found to {\it not} mix completely
(e.g.\ White 1980, Barnes 1988), 
and a wide variety of observational evidence, such as the 
presence of metallicity gradients in elliptical galaxies, 
supports the idea that the most bound material in the progenitors
tends to remain the most bound in the merger product.
%White, S.D.M. 1987, in Structure and Dynamics of Elliptical
%galaxies, ed. T de Zeeuw (Dordrecht: Reidel), p.\ 339.
It is also important to note that this thorough 
mixing must continue until recent epochs. If a population
was thoroughly mixed in galactic radius at early times, but was mostly
in place at a redshift of one, a radial dependence of the
GCLF would be created, as much of the evolution of the cluster
would be dominated by its tidal radius at its current location
(see Figure 5 in Vesperini et al.\ 2003).

Models 2 and 3 which introduce physical processes other than
two-body relaxation to help set the scale of the turnover of the
globular cluster mass function naturally avoid this problem of
the radial constancy of the GCLF. They do not require strongly
radial orbits to achieve a constant GCLF, and thus do not
violate the observed anisotropy constraints. By far the biggest
challenge to these models is the evidence that the mass function
for young globular cluster systems is a power-law extending to
masses below the current turnover mass observed in old
systems. This power-law to lower masses is best established
for the Antennae (Zhang \& Fall 1999), but is also consistent
with a wide range of other data on young cluster systems.
The key question is whether the age of the bulk of the Antennae
clusters is older than the age by which these early processes
will have disrupted low-mass clusters. As Parmentier \& Gilmore (2007)
and others point out, disruption by gas expulsion is not absolutely 
immediate, but would be expected to take of order of tens of Myr
for an unbound cluster to disperse itself into the field. The 
bulk of the Anntenae cluster sample ranges up to about 100 Myr, 
so this is somewhat older, although it can be argued whether this
difference is large enough to be fatal for the model. 

The hypothesis that the low-mass clusters are disrupted somewhat
later, due to stellar mass loss after they are purely stellar
systems (Vesperini \& Zepf 2003), has an important advantage 
in comparison with the Antennae data that it happens later than 
the proposed gas expulsion. One key here is establishing whether 
the globular cluster mass-concentration relation observed for the
Milky Way globular clusters is primordial. If it is, it seems
unavoidable that stellar mass loss will cause some preferential
dissolution of low-mass globular clusters, and cause the
globular cluster mass function to flatten or possibly turnover
at lower masses
(Vesperini \& Zepf 2003 and discussion therein).
Another key test of the mechanism by which most low-mass globular
clusters are destroyed is to study intermediate-aged globular
cluster systems, as there is no doubt these are old enough any
early destruction mechanism should have changed their mass functions.
To date this has proved challenging. Although deep optical data alone
may not be able to distinguish intermediate and old globular
cluster systems, results suggestive of a power-law mass function 
below the turnoff mass in intermediate-age systems have been 
found by Paul Goudfrooij and collaborators (e.g.\ Goudfrooij et al.\ 2007). 
One way forward is deep HST near-infrared imaging
like that obtained for the intermediate-age system in NGC~4365
(Kundu et al.\ 2005) which allows for a clean enough separation
of globular cluster sub-populations to test the dependence
of the mass function on age.

\section{The Implications of the Weak Mass-Radius Relationship for Globular}

The masses and radii of globular clusters are for the most part
uncorrelated, with only a very shallow relation between the two.
This has been established both in the Milky Way (e.g.\ van den 
Bergh et al.\ 1991, Djorgovski \& Meylan 1994, Ashman \& Zepf 1998), 
and for extragalactic cluster systems (e.g.\ Waters et al.\ 2006,
Jord\'an et al.\ 2005).
This absent or very shallow mass-radius relation stands in
contrast to nearby every other type of astronomical object
to which globular clusters might be compared. For example,
galaxies have a clear mass-radius relation, and clusters 
of galaxies do as well. Perhaps most importantly scaling relations for
molecular clouds, the natural progenitors of star clusters,
give $R \propto M^{0.5}$. Studies of globular clusters,
both Galactic and extragalactic, give a dramatically
shallower relation, with the best current constraint
probably being the $R \propto M^{0.04}$ found for M87 globulars
by Waters et al.\ (2006).

Furthermore, studies of young globular cluster systems have also
found a very weak mass-radius relation, first in NGC~3256 (Zepf et al.\ 1999),
and now in a number of galaxies (Larsen 2004, Scheepmaker et al.\ 2007).
The lack of a mass-radius correlation in both young and old
globular cluster systems strongly suggests the reason for its
surprising absence is {\it not} due to a long-term evolutionary
process, but must be closely related to the formation and early 
evolution of globular clusters.

The question then is how the formation and early evolution
of globular cluster produces such a weak mass-radius relation,
particularly when the progenitor clouds seem to have a typically
strong relation. Ashman \& Zepf (2001) considered many possibilities
for physical mechanisms to account for the weak mass-radius relation
of globular clusters. Most of these failed, including such
standard ideas as a Schmidt law relating star formation
efficiency to density. The one solution that works is to
adopt a star formation efficiency proportional to the binding
energy of the molecular cloud. Because lower mass clusters
have less binding energy per unit mass than higher mass
clusters, in this case, low mass clusters have a lower star
formation efficiency. As a result of this lower star formation
eficiency, when the remaining gas is lost from the clusters,
low mass clusters will expand more in response to this mass
loss. 

Note that this is generally true of any proposal in which
lower-mass young clusters lose a greater fraction of their
mass than higher mass clusters. Lower mass clusters then respond
to this greater mass loss by expanding more than higher mass
clusters, assuming the mass loss happens adiabatically. Thus,
the radii of low and high-mass clusters become more similar.
The same effect might occur in cases in which the
cluster loses mass because of stellar mass loss, if more massive
globular clusters lose a smaller fraction of their mass.

An invariable outcome of models that successfully produce
a weak mass-radius relation is that they flatten the globular
cluster mass function (Ashman \& Zepf 2001). Exactly how much 
they do so depends on the specifics of the model, both 
how the expansion of a cluster is related to the star
formation efficiency, and what the final mass-radius relation
is. For example, adopting the final radius of
the cluster $r_f$ is the initial radius $r_i$ divided by
the efficiency $\epsilon$, that is $r_f = r_{i}/\epsilon $,
and the efficiency is proportional to the binding energy
per unit mass of the initial system to the power $n$,
that is $\epsilon \propto (M_{i}/r_{i})^{n}$, then the
final slope of the globular cluster system mass function, $\alpha$
is $\alpha = (2\beta +n)/(n+2)$, where 
$\beta$ to be the initial slope  
(see Section 4 of Ashman \& Zepf 2001). If $n$ is around 0.5,
corresponding to a significant but modest mass-radius relation,
then the difference between the initial and final mass slopes
is small, about $10\%$. If $n$ is large, such as $n=1$ which
gives no mass-radius relation, then the difference between
the initial and final mass slopes is larger.

Real cluster evolution is undoubtedly more complicated,
but these calculations provide an effective framework for
showing the connections between the question of the
weak mass-radius relation and of the globular cluster
mass function. Specifically, the most viable current
proposal for the origin of the weakness of the mass-radius
relation for globular clusters generically produces a
change a flattening of the globular cluster mass function
(see Ashman \& Zepf 2001). Whether this flattening
is substantial or modest depends on the exact mass-radius
relation and to some extent on whether the mass loss and
expansion occurs adiabatically. Therefore, one of the
obvious observational challenges is to determine the
mass-radius relation for globular cluster systems as
accurately as possible. A second challenge is to determine
the mass function in globular cluster systems that are
old enough to have experienced the bulk of their stellar
mass loss. If these still have steep power law mass functions 
like those of molecular clouds, then either the mass-radius
relation must be pushed to the limits of the current
constraints, or another solution for the weakness of
the mass radius relation that does not effect the
cluster mass function must be found. The underlying
key points are that the data for both young and old globular 
cluster systems indicate a very weak mass-radius relation,
and that extant explanations to produce this have
implications for the globular cluster mass function.

\begin{acknowledgments}

Much of the work described above has been carried out in
collaboration with Chris Waters, Enrico Vesperini, and Keith
Ashman. I gratefully acknowledge support for this work from 
NSF award AST-0406891 and grants GO-8592 and GO-10543 from the Space
Telescope Science Institute.

\end{acknowledgments}

%\begin{discussion}
%\end{discussion}

\end{document}